\documentstyle[12pt]{article}
\title{\bf An action of the gluodynamics string from the 
Wilson loop expansion}
\author{D.V.ANTONOV \thanks{E-mail:
antonov@pha2.physik.hu-berlin.de, antonov@vxitep.itep.ru.~~ 
Supported by the Russian Fundamental Research Foundation, Grant No.
96-02-19184 and by the INTAS, Grant No.94-2851.}
\\
{\it Institute of Theoretical and Experimental Physics,}\\
{\it B.Cheremushkinskaya, 25, 117 218, Moscow, Russia}\\
{\it and}\\
{\it Institut f\"ur Elementarteilchenphysik,
Humboldt-Universit\"at,}\\ 
{\it Invalidenstrasse 110, D-10115, Berlin, FRG}\\
\\
D.EBERT \thanks{E-mail: debert@qft2.physik.hu-berlin.de}\\
{\it Institut f\"ur Elementarteilchenphysik,
Humboldt-Universit\"at,}\\
{\it Invalidenstrasse 110, D-10115, Berlin, FRG}}
\date{}
\begin{document}
\maketitle
\vspace{1cm}

\newcommand{\be}{\begin{equation}}
\newcommand{\ee}{\end{equation}}

\vspace{1cm}
\centerline{\bf {Abstract}}

\vspace{3mm}
The nonabelian Stokes theorem, representing a Wilson loop as an
integral over all the orientations in colour space, and the cumulant
expansion are used for derivation of the string effective action in
$SU(2)$ gluodynamics. The obtained theory is the theory of the rigid
string interacting with the rank two antisymmetric Kalb-Ramond
fields. In this model there exists a phase where there are no problems
of crumpling and wrong high temperature
behaviour of the string tension, which are present in the free rigid
string theory. The Langevin approach to stochastic quantization of
the obtained theory is applied.

\newpage
{\large \bf 1. Introduction}

\vspace{3mm}
There exist several arguments why the Nambu-Goto string is a bad
candidate for the gluody-\\
namics string. First it contradicts with the
parton-like behaviour observed in deep inelastic scattering, since 
the observed scattering amplitudes have a power law fall-off, while
the Nambu-Goto string gives rise to the exponential fall-off. Second
an averaged Wilson loop $<W(C)>$ is invariant under
orientation-preserving reparametrizations of the oriented contour $C$:
$<W(C(s))>=<W(C(\alpha(s))>$ when $\alpha^{\prime}(s)>0$, while the
Nambu-Goto action is not sensitive to the sign of the
reparametrization and therefore cannot be a string action
corresponding to the Wilson loop expansion. Another fact of
fundamental importance is that the gluodynamics string is an
asymptotical theory, which exists only at the distances larger than
the correlation length of the vacuum $T_g~^1$. This property can be
taken into account if one adds to the Nambu-Goto term the so-called
rigidity term$^{2,3}$, and it was shown in$^{4}$ that the rigidity
term actually appears in the expansion of the averaged Wilson loop in
powers of $\frac{T_g}{r}$, where $r$ is the size of the Wilson loop.
Such an expansion is valid in the confining regime of the Wilson loop,
when the ratio $\frac{T_g}{r}$ is small$^{5}$. This expansion allows
one to establish the criterion of confinement in terms of the two
scalar functions, which parametrize the bilocal correlator of gluonic
fields$^{6}$.

However there are some obstacles which destroy the status of the rigid
string as the gluodynamics string. The most serious of them is that the
operator product expansion in the rigid string theory contains a local
operator of dimenion two$^{7}$, while the operator product expansion
in gluodynamics cannot involve any operators of dimension less than
four. Second as a higher-derivative theory the rigid string theory
has no the lowest-energy state$^{8}$. Third if the $\beta$-function
has no zeros the string world
sheet is creased (normals are short-ranged), which leads to the so-called 
problem of crumpling$^{2,9}$ (for a 
review see$^{10}$). In this case there presents a tachyon (the
state with a negative norm) in the string spectrum$^{2,9}$.
The last problem of the rigid string
theory is that at high temperatures the value of the free energy per
unit length (string tension) does not coincide with the behaviour
derived from gluodynamics$^{11}$. The problems of nonexistence of a 
lowest-energy state, presence of a tachyon in the spectrum, and the
wrong high temperature behaviour of the string tension are absent in
the new nonlocal theory of a string with
the negative stiffness, which was proposed in$^{12}$. However it
remains unclear how
this model can be derived from the gluodynamics partition function or
from the Wilson loop expansion.

An alternative approach to the gluodynamics string was suggested
in$^{13}$, where it was shown that a phase transition occured in the
rigid string theory coupled to the Kalb-Ramond fields$^{14}$, so that
the new phase was characterized by the long range order of normals
and therefore absence of crumpling. Moreover in$^{15}$ it was
demonstrated that this model gave a consistent solution for the free
energy of gluodynamics at high temperatures and thus cured the
corresponding problem of the free rigid string theory.

The aim of this letter is to show how the theory of a rigid string,
interacting with the Kalb-Ramond fields, may be derived via the
expansion of the Wilson loop written through the nonabelian Stokes
theorem and the cumulant expansion$^{6,16,20}$. There exist in
literature two versions of the nonabelian Stokes theorem: first was
suggested in$^{16,17}$, and its important property is the presence of
the path ordering, while in the second version, suggested in$^{18}$,
the path ordering was replaced by the integration over an auxiliary
scalar field from the $SU(N_c)/[U(1)]^{(N_c-1)}$ coset space. For the
purposes stated above it occures crucial to use the second version of
the nonabelian Stokes theorem.

After that we shall apply the Langevin approach$^{19}$ to stochastic
quantization of the obtained model. Using the Fourier transformation
we shall find the retarded Green functions of the diffe- \\
rential
operators standing in the Langevin equations and reduce these
equations to the integral-differential ones, which can be solved
perturbatively.

The
main results of the letter are summarized in the Conclusion.

\vspace{6mm}
{\large \bf 2. Rigid string coupled to the Kalb-Ramond fields from the
Wilson loop expansion and the Langevin approach to its quantization}

\vspace{3mm}
We shall start with the nonabelian Stokes theorem, suggested
in$^{18}$, and consider the $SU(2)$ case. Then one can write the
Wilson loop $W(C)=tr P exp\left(i\oint\limits_C^{} A_\mu^a t^a
dx_\mu\right)$ in the following way$^{18}$

$$W(C)=\int D \vec n~ exp~ \Biggl[ \frac{iJ}{2} \Biggl(-\int\limits_S^{}
d\sigma_{\alpha\beta} F_{\alpha\beta}^i n^i + \int\limits_S^{} 
d\sigma_{\alpha\beta} \varepsilon^{ijk} n^i (D_\alpha \vec
n)^j (D_\beta \vec n)^k \Biggr) \Biggr], \eqno (1)$$
where $\vec n$ is a unit 3-vector which characterizes the instant
orientation in the colour space, $J=\frac{1}{2},$\\
$1, \frac{3}{2}, ...$ 
is the spin of the representation of the Wilson loop considered,
$F_{\alpha \beta}^i =\partial_\alpha A_\beta^i-\partial_\beta
A_\alpha^i+$\\
$+\varepsilon^{ijk}A_\alpha^j A_\beta^k,
D_\alpha^{ij}=\delta^{ij}\partial_\alpha
-\varepsilon^{ijk}A_\alpha^k$, and $S$ is any surface bounded by the
contour $C$. It should be emphasized that the representation of the
Wilson loop is fixed to be exactly $J$ by the second term on the right
hand side of equation (1) (the so-called Wess-Zumino term), which
breaks down the isotropy of the $\vec n$-space. Therefore while
dealing with the Wilson loop in a given representation, considering
the integration over $\vec n$ in (1) as some averaging procerure and
using the cumulant expansion$^{6,16,20}$ one gets in the bilocal
approximation$^{6}$

$$W(C)=exp \Biggl[-\frac{iJ}{2}\int\limits_S^{}
d\sigma_{\alpha\beta}<G_{\alpha\beta}>_{\vec n}-
\frac{J^2}{4}\int\limits_S^{} 
d\sigma_{\alpha \beta}(x) d\sigma_{\mu\nu}(x^\prime) \ll G_{\alpha
\beta}(x) G_{\mu\nu}(x^\prime) \gg_{\vec n} \Biggr], \eqno (2)$$
where $G_{\alpha\beta}=F_{\alpha\beta}^i n^i-\varepsilon^{ijk}n^i (D_
\alpha \vec n)^j(D_\beta \vec n)^k$, so that the first term on the
right hand side of equation (2) does not vanish (since the averaging
is performed only over the $\vec n$-field, but not over the physical
vacuum), and $<G_{\alpha\beta}>_{\vec n}$ is some antisymmetric tensor
field. 

In what follows we shall use the method suggested in$^{4}$ and
consider $-ln W(C)$ with $W(C)$ defined via (2) as a gluodynamics
string effective action.
At this point we shall quote its resulting form (5), which is
the action of the rigid string interacting with the Kalb-Ramond
fields. Gaussian integration over the latter in (5) yields the long
range Coulomb potential$^{13}$

$$ V(x(\xi)-x(\xi^\prime),a)=\frac{2g_0}{\pi}
\frac{1}{(x(\xi)-x(\xi^\prime))^2 +a^2\sqrt{g}},$$
with the coupling constant $g_0=\alpha_0\alpha_{Coulomb}\equiv
\alpha_0 \frac{e^2}{4\pi}$, where in order to avoid the singularity at
$\xi= \xi^\prime$ we introduced a cut-off $a$, which was reasonable to
be taken of the order of the correlation length of the vacuum $T_g$.
Let us then introduce a dimensionless field $B_\mu^i=a A_\mu^i$
and an auxiliary Abelian field $H_\mu$, which satisfies the equation

$$\gamma\int d^4x
\biggl(\varepsilon_{\mu\nu\lambda\rho}{\cal
F}_{\mu\nu}\partial_\lambda H_\rho
- \gamma H_\mu^2\biggr)= iJ\int\limits_S^{}
d \sigma_{\alpha\beta} \varepsilon^{ijk} <n^i (D_\alpha \vec n)^j(D_\beta
\vec n)^k>_{\vec n},\eqno (3)$$
where $\gamma$ is an arbitrary parameter, ${\cal F}_{\mu\nu}=<f_{\mu\nu}^i
n^i>_{\vec n}, f_{\mu\nu}^i=\partial_\mu B_\nu^i-
\partial_\nu
B_\mu^i
+\frac{1}{a}\varepsilon^{ijk}B_\mu^jB_\nu^k$. The dependence
on the parameter $\gamma$ drops out when one integrates over the field
$H_\mu$, which leads to the Lagrangian of the antisymmetric tensor
field

$${\cal L}=-\frac{1}{12} K_{\mu\nu\lambda}K_{\mu\nu\lambda}, \eqno
(4)$$
where $K_{\mu\nu\lambda}=\partial_\mu{\cal
F}_{\nu\lambda}+\partial_\nu {\cal F}_{\lambda\mu}+\partial_\lambda
{\cal F}_{\mu\nu}$. Parametrizing the bilocal cumulant $\ll
G_{\alpha\beta} (x)G_{\mu\nu} (x^\prime)\gg_{\vec n}$ in the following
way$^{6,16}$

$$\ll G_{\alpha\beta}(x) G_{\mu\nu}(x^\prime)\gg_{\vec n}=
(\delta_{\alpha \mu} \delta_{\beta\nu}-\delta_{\alpha\nu}
\delta_{\beta \mu}) D\Biggl(\frac{(x-x^\prime)^2}{T_g^2}\Biggr)+
\frac{1}{2} \Biggl[\frac{\partial}{\partial x_\alpha} ((x-x^\prime)_\mu
\delta_{\beta\nu} -(x-x^\prime)_\nu\delta_{\beta\mu})+ $$

$$+\frac{\partial}{\partial x_\beta}((x-x^\prime)_\nu
\delta_{\alpha\mu}-
(x-x^\prime)_\mu
\delta_{\alpha\nu})\Biggr]D_1\Biggl(\frac{(x-x^\prime)^2}{T_g^2}
\Biggr),$$
where $D$ and $D_1$ are two coefficient functions, expanding the
second term on the right hand side of (2) in powers of
$\frac{T_g}{r}$, where $r$ is the size of the Wilson loop, in the same
manner as it was done in$^{4}$ and using (3) and (4), we finally get
from equation (2) the following action of the $SU(2)$ gluodynamics
string in the bilocal approximation 

$$S_{biloc.}=\sigma\int d^2\xi \sqrt{g} +\frac{1}{2\alpha_0}\int d^2\xi
\Biggl[\frac{1}{\sqrt{g}}
(\partial^2x)^2+\lambda^{ab}((\partial_ax_\mu)(\partial_b
x_\mu)-\sqrt{g}\delta_{ab}) \Biggr]+ $$

$$+e_0\int d^2\xi\varepsilon^{ab}(\partial_a x_\mu)(\partial_b
x_\nu)\phi_{\mu \nu}+\frac{1}{12}\int d^4y
P_{\mu\nu\lambda}P_{\mu\nu\lambda}+O\Biggl[
max\Biggl(\frac{T_g^6D(0)}{r^2}, \frac{T_g^6D_1(0)}{r^2}\Biggr)
\Biggr]. \eqno (5)$$
In (5) $\sigma=J^2T_g^2\int d^2z D(z^2), \alpha_0=\frac{16}{J^2T_g^4
\int d^2z z^2(2D_1(z^2)-D(z^2))}, e_0=\frac{J}{2a}, \lambda^{ab}$ is
the Lagrange multiplier, $\phi_{\mu\nu}=i{\cal F}_{\mu\nu}, 
P_{\mu\nu\lambda}=iK_{\mu\nu\lambda}, 
g=det\parallel g_{ab} \parallel$, and we have used the conformal gauge
$g_{ab}=\sqrt{g}\delta_{ab}$. 
One can now see that all the
dependence on the spin of the representation of the Wilson loop has
gone into the coupling constants.

Thus expanding the Wilson loop, written through the nonabelian
Stokes theorem suggested in$^{18}$, we obtained an action of the
$SU(2)$ gluodynamics string, which occured to be the action of the
rigid string coupled to the rank two antisymmetric Kalb-Ramond field
$\phi_{\mu \nu}$$^{14}$. It was shown in$^{13}$ that there existed two
phases in this theory, which were distinguished by the values of the
order parameter $m$, defined from the mass gap equation

$$\int d^2p \frac{p^2}{p^2(p^2+m^2)+
p^2 V_0(p)+V_1(p)}=\frac{2 \pi^2}{\alpha_0},$$ 
where $V_0(p)=8g_0K_0(a|p|),
V_1(p)= \frac{8g_0}{a^2}(a|p|K_1(a|p|)-1), K_0$ and $K_1$ were the Macdonald
functions. In the disordered phase, corresponding to the values $m>0$,
the coupling constants $\alpha_0$ and $g_0$ are fixed by dimensional
transmutation in terms of $m$ and the cut-off $a$, while the phase
corresponding to $m=0$ is characterized by the long range order of
normals (absence of crumpling) and therefore may describe the
gluodynamics string, as it could be expected. Notice also that
in$^{21}$ the renormalized mass gap equation was derived, and it was
shown that the phase transition survived quantum fluctuations.

To conclude with we shall apply the Langevin approach$^{19}$
to stochastic
quantization of our theory.
To this end, by making use of the conformal gauge, we shall rewrite
the Nambu-Goto term on the right hand side of equation (5) in an
equivalent form$^{22}$ $\frac{\sigma}{2}\int d^2 \xi (\partial^a
x_\mu) (\partial_a x_\mu)$ and then neglect for simplicity the
rigidity term, since it is of the highest order in $T_g$ in comparison
with the others. 
Therefore one gets
from (5) the following Langevin equations

$$\dot x_\mu-\sigma\partial^a\partial_a
x_\mu= \eta_\mu-\frac{e_0}{3}P_{\mu\nu\lambda}
\varepsilon^{ab}(\partial_a
x_\nu)(\partial_b x_\lambda), \eqno (6)$$

$$\dot \phi_{\mu\nu}-(\partial_\alpha\partial_\alpha \phi_{\mu\nu}+
\partial_\mu
\partial_\lambda\phi_{\nu\lambda}+\partial_\nu\partial_\lambda
\phi_{\lambda\mu})=\eta_{\mu\nu} -e_0\delta^{(2)}(\xi_\perp)
\varepsilon^{ab}(\partial_ax_\mu)(\partial_b
x_\nu), \eqno (7)$$
where in (6) the Langevin time $t$ has the dimension of the fourth
power of length while in (7) it has the dimension  of the square of
length, $\eta_\mu$ and $\eta_{\mu\nu}$ are two Gaussian noises, whose
bilocal correlation functions have the form
$<\eta_\mu(x,t)\eta_\nu(x^\prime,
t^\prime)>= 2\delta_{\mu\nu}\delta(x-x^\prime)\delta(t-t^\prime),
<\eta_{\mu\nu}(x,t)\eta_{\alpha\beta}(x^\prime,
t^\prime)>=$ \\
$=2(\delta_{\mu\alpha}
\delta_{\nu\beta}-\delta_{\mu\beta}\delta_{\nu\alpha})
\delta(x-x^\prime)\delta(t-t^\prime) $, and $\xi_\perp^a$ is a
hyperplane perpendicular to the hyperplane $\xi^a$.

The retarded Green functions of the operators standing on the left
hand sides
of equations (6) and (7) can be 
obtained via the Fourier transformation, and one gets from (6) and (7)
the following integral-differential equations

$$x_\mu(\xi,t)=\int\limits_0^t dt^\prime\int d^2\xi^\prime
G(\xi-\xi^\prime, t-t^\prime)\Biggl[\eta_\mu(\xi^\prime,
t^\prime)-\frac{e_0}{3} P_{\mu\nu\lambda}(x(\xi^\prime,t^\prime
), t^\prime)
\varepsilon^{ab} \Biggl(\frac{\partial}{\partial \xi^{\prime a}}
x_\nu(\xi^\prime ,
t^\prime)\Biggr)\cdot $$

$$\cdot \Biggl(\frac{\partial}{\partial\xi^{\prime
b}}x_\lambda(\xi^\prime, t^\prime)\Biggr)\Biggr], \eqno (8)$$

$$\phi_{\mu\nu}(x,t)=\int\limits_0^t dt^\prime \int d^4 x^\prime
{\cal G}_{\mu\nu, \alpha\beta} (x-x^\prime, t-t^\prime)\biggl[
\eta_{\mu\nu}( x^\prime, t^\prime)-e_0\delta^{(2)}(\xi_\perp)
\varepsilon^{ab} (\partial_a x_\mu^\prime)(\partial_b x_\nu^\prime)
\biggr], \eqno (9)$$
where the Green functions $G(\xi,t)$ and ${\cal G}_{\mu\nu,
\alpha\beta} (x,t)$ read

$$G(\xi,t)=\frac{e^{-\frac{\xi^2}{4\sigma t}}}{4\pi \sigma t}, $$

$${\cal G}_{\mu\nu, \alpha\beta}(x,t)=\frac{1}{96}\Biggl[ (\delta_{\mu\alpha}
\delta_{\nu\beta} -\delta_{\mu\beta}\delta_{\nu\alpha})
\Biggl(\frac{e^{- \frac{x^2}{4t}}}{\pi^2 t^2}+32\delta (x)\Biggr)+
\frac{1}{x^2} \biggl(\delta_{\mu\alpha}x_\nu x_\beta +
\delta_{\nu\beta}x_\mu x_\alpha -\delta_{\mu\beta}x_\nu x_\alpha
-\delta_{\nu\alpha} x_\mu
x_\beta\biggr)\cdot $$

$$\cdot \Biggl(\frac{e^{-\frac{x^2}{4t}}}{\pi^2 t^2}-16\delta
(x) \Biggr)\Biggr], $$
and the initial conditions $x_\mu (\xi,0)=0, \phi_{\mu\nu}(x,0)=0$
were implied. Solving equations (8) and (9) perturbatively one can
develop stochastic diagrammatic technique in the model under
consideration
in the coordinate
representation.
 
\vspace{6mm}
{\large \bf 3. Conclusion}

\vspace{3mm}
In this letter we have shown how the Wilson loop expansion in $SU(2)$
gluodynamics generates \\
string effective action. To this end we have
written the Wilson loop via the nonabelian Stokes \\
theorem suggested
in$^{18}$ and applied to it the cumulant expansion$^{6,16,20}$ in the
bilocal approximation$^{6}$. After that, introducing an auxiliary
Abelian field, which satisfies integral-differential equation (3), and
integrating it over, we eliminated the coupling of the rigidity string
term with the Wess-Zumino term, which fixed the representation of the
Wilson loop. Then, Taylor expanding the second term of the cumulant
expansion in powers of $\frac{T_g}{r}$, where $T_g$ is the correlation
length of the vacuum$^{5,6}$, and $r$ is the size of the Wilson loop,
we arrived at the action of the rigid string interacting with the rank
two antisymmetric Kalb-Ramond tensor fields. This interaction is the
long ranged Coulomb interaction, which should be cut off at the
distances of the order of $T_g$, and its coupling constant is
proportional to the spin of the representation of the Wilson loop and
inversly proportional to the Coulomb cut-off. The dependence on the
spin of the representation also appears in the Nambu-Goto string
tension and the coupling constant of the rigidity term in contrast
to$^{4}$, where another version of the nonabelian Stokes
theorem$^{16,17}$ was applied to derivation of the string effective
action. 

The obtained theory provides correspondence between its coupling
constants and the gluodynamics coupling constant at high
temperatures$^{15}$ in contrast to the free rigid string theory$^{11}$
and possesses a phase, where the normals to the string world sheet are
long ranged$^{13}$ and therefore is a much better candidate for the
gluodynamics string than the free rigid string.

Finally we derived the Langevin equations
in this theory, where the highest in $T_g$ rigidity term was
neglected. After that we found the retarded Green functions of the
differential operators, standing in these equations, and reduced the
problem to the system of two integral-differential equations, which
can be solved perturbatively.

\vspace{6mm}
{\large \bf 4. Acknowledgements}

\vspace{3mm}
D.A. is deeply grateful to Professors Yu.M.Makeenko and Yu.A.Simonov
for useful discussions and to the theory group of the Quantum Field
Theory Department of the Institut f\"ur Physik of the
Humboldt-Universit\"at of Berlin for kind hospitality.

\newpage
{\large\bf References}

\vspace{3mm}
\noindent
1.~ Yu.A.Simonov, ITEP-28-92, {\it Nuovo Cim.} {\bf A107}, 2629 (1994).\\
2.~ A.M.Polyakov, {\it Nucl.Phys.} {\bf B268}, 406 (1986).\\
3.~ H.Kleinert, {\it Phys.Lett.} {\bf B174}, 335 (1986); 
T.L.Curtright et al., {\it Phys.Rev.} {\bf D34}, 3811 (1986);
G.Germ\'an, {\it Mod.Phys.Lett.}
{\bf A6}, 1815 (1991).\\
4.~ D.V.Antonov et al., {\it Mod.Phys.Lett.} {\bf A11}, 1905 (1996)
(DESY-96-134).\\
5.~ M.Campostrini et al., {\it Z.Phys.} {\bf C25}, 173 (1984);
I.J.Ford et al., {\it Phys.Lett.} {\bf B208}, 286 (1988); A. Di
Giacomo and H. Panagopoulos, {\it Phys.Lett.} {\bf B285}, 133 (1992);
E.Laermann et al., {\it Nucl.Phys.} {\bf B26} (Proc.Suppl.), 268
(1992).\\
6.~ H.G.Dosch, {\it Phys.Lett.} {\bf B190}, 177 (1987); Yu.A.Simonov,
{\it Nucl.Phys.} {\bf B307}, 512 (1988); H.G.Dosch and Yu.A.Simonov,
{\it Phys.Lett.} {\bf B205}, 339 (1988), {\it Z.Phys.} {\bf C45}, 147
(1989); Yu.A.Simonov, {\it Nucl.Phys.} {\bf B324}, 67 (1989), {\it
Phys.Lett.} {\bf B226}, 151 (1989), {\it Phys.Lett.} {\bf B228}, 413
(1989); for a review see Yu.A.Simonov, {\it Yad.Fiz.} {\bf 54}, 192
(1991).\\
7.~ A.M.Polyakov, unpublished.\\
8.~ S.W.Hawking, in {\it Quantum Field Theory and Quantum Statistics},
eds. I.A.Batalin et al. (Hilger, 1987); E.Braaten and C.K.Zachos, {\it
Phys.Rev.} {\bf D35}, 1512 (1987).\\
9.~ A.M.Polyakov, {\it Gauge Fields and Strings} (Harwood, 1987).\\
10. F.David, {\it Phys.Rep.} {\bf 184}, 221 (1989).\\
11. J.Polchinski and Z.Yang, {\it Phys.Rev.} {\bf D46}, 3667 (1992).\\
12. H.Kleinert and A.M.Chervyakov, {\it hep-th} /9601030.\\
13. M.Awada and D.Zoller, {\it Phys.Lett.} {\bf B325}, 115 (1994).\\
14. M.Kalb and P.Ramond, {\it Phys.Rev.} {\bf D9}, 2273 (1974).\\
15. M.Awada, {\it Phys.Lett.} {\bf B367}, 270 (1996).\\
16. Yu.A.Simonov, {\it Yad.Fiz.} {\bf 50}, 213 (1989).\\
17. M.B.Halpern, {\it Phys.Rev.} {\bf D19}, 517 (1979); I.Ya.Aref'eva,
{\it Theor.Math.Phys.} {\bf 43}, 111 (1980).\\
18. D.Diakonov and V.Petrov, in {\it Nonperturbative approaches to
QCD}, Proceedings of the International Workshop at ECT*, Trento, July
10-29, 1995, ed. D.Diakonov (PNPI, 1995).\\
19. G.Parisi and Y.-S.Wu, {\it Sci.Sin.} {\bf 24}, 483 (1981).\\
20. N.G.Van Kampen, {\it Stochastic Processes in Physics and
Chemistry} (North-Holland Physics Publishing, 1984).\\
21. M.Awada, {\it Phys.Lett.} {\bf B351}, 468 (1995).\\
22. M.Green, J.Schwarz, E.Witten, {\it Superstring Theory} (Cambridge
University Press, 1987).\\
\end{document}